\documentclass[aps, prd, amsmath, floats, floatfix, twocolumn, nofootinbib, showpacs]{revtex4-1}
\usepackage{graphicx}
\usepackage{color}

\def\beq{\begin{equation}}
\def\eeq{\end{equation}}
\def\beqn{\begin{eqnarray}}
\def\eeqn{\end{eqnarray}}

\newcommand{\be}{\begin{eqnarray}}
\newcommand{\ee}{\end{eqnarray}}

\begin{document}

\title{K$\alpha$ iron line profile from accretion disks around regular and singular exotic compact objects}

\author{Cosimo Bambi}
\email{bambi@fudan.edu.cn}

\author{Daniele Malafarina}
\email{daniele@fudan.edu.cn}

\affiliation{Center for Field Theory and Particle Physics \& Department of Physics, Fudan University, 200433 Shanghai, China}

\date{\today}

\begin{abstract}
The nature of the super-massive black hole candidates in galactic nuclei can 
be tested by analyzing the profile of the K$\alpha$ iron line observed in their 
X-ray spectrum. In this paper, we consider the possibility that the spacetime
in the immediate vicinity of these objects may be described by some 
non vacuum exact solutions of Einstein's equations resulting as the 
end-state of gravitational collapse. The vacuum far away portion of
the spacetime is described by the Schwarzschild metric, while the interior 
part may be either regular or have a naked singularity at the center. 
The iron line generated around this class of objects has specific features, 
which can be used to distinguish such objects from Kerr black holes. 
In particular, their iron line cannot have the characteristic low-energy 
tail of the line generated from accretion disks around fast-rotating 
Kerr black holes. We can thus conclude that the super-massive black 
hole candidates whose spin parameter has been estimated to be close 
to 1 assuming the Kerr background cannot be this kind of objects.  
\end{abstract}

\pacs{04.20.Dw, 04.20.Jb, 98.62.Js}

\maketitle


\section{Introduction}

Astrophysical black hole (BH) candidates are stellar-mass compact
objects in X-ray binary systems and super-massive dark bodies at the
center of every normal galaxy. They are thought to be the Kerr BH of 
general relativity, but their actual nature has still to be verified~\cite{review}.
In fact, the only robust measurements we have so far are for the masses of these objects. 
In the case of stellar-mass BH candidates, the mass exceeds 3~$M_\odot$, 
which is the maximum value for the mass of a neutron star for any reasonable 
matter equation of state~\cite{bh1}. Super-massive BH candidates in 
galactic nuclei are simply too heavy, compact, and old to be clusters of 
non-luminous objects, as the cluster lifetime due to evaporation and physical 
collisions would be shorter than the age of these systems~\cite{bh2}. 
The non-observation of electromagnetic radiation emitted by the possible
surface of these objects may also be interpreted as an evidence for the
existence of an event horizon~\cite{horizon}, which is the key-feature of
BHs (see however~\cite{horizon2} and note that long-living trapped 
surfaces can mimic an event horizon~\cite{horizon3}). The interpretation 
that astrophysical BH candidates are the Kerr BHs of general relativity is 
therefore the most natural explanation, and the only one that does not 
explicitly require new physics. However, an observational confirmation of this 
hypothesis would be at least desirable. Attempts to test the actual nature of 
these objects can be found in~\cite{cb,cbc,kss,torres,harko,jp,others}.

In 4-dimensional general relativity, BHs are described by the Kerr solution
and they are completely specified by two parameters: the mass $M$ and
the spin angular momentum $J$. However, there are many solutions that 
are not BHs and can describe the gravitational field outside a super-massive 
compact object. These can vary from non-spherical vacuum solutions with 
naked singularities to solutions with a non-vanishing energy momentum 
tensor. A super-massive compact object described by a so called ``interior'' 
solution that matches smoothly to a known vacuum exterior could in principle 
be an alternative explanation to the BH paradigm. In the present paper, we
explore such a possibility. In~\cite{dm}, one
of us studied the thermal spectrum of geometrically thin disks around such
kind of sources, in order to determine how they would be observationally 
different from a BH. The non-existence of an innermost stable circular orbit
(ISCO) in this class of metrics is the key-point to understand their spectrum.
The radiative efficiency turns out to be very high and therefore the thermal
spectrum of the disk is harder than the one around a Kerr BH with spin
parameter $a_* = J/M^2 = 1$. So, the stellar-mass BH candidates in X-ray
binary systems cannot be this kind of objects. In the case of the super-massive 
BH candidates in galactic nuclei, we cannot observe the thermal spectrum 
of their thin disks (the disk's temperature is proportional to $M^{-0.25}$ and
for $M \sim 10^6 - 10^9$~$M_\odot$ the spectrum is in the UV range, where 
dust absorption makes measurements impossible). The analysis of the
K$\alpha$ iron line is currently the only available technique to probe the
spacetime geometry around super-massive BH candidates. The present 
work is therefore devoted to the study of the iron line profile of these exotic
compact objects to find observational features to test the possibility that they
can be the super-massive BH candidates at the center of galaxies.

The content of the paper is as follows. In Section~\ref{s-ka}, we briefly review
the origin and the features of the K$\alpha$ iron line profile and its role to
test the nature of BH candidates. In Section~\ref{s-sol}, we introduce the
interior solutions of the exotic compact objects we are going to investigate.
The profiles of the K$\alpha$ iron line from this class of sources are calculated
in Section~\ref{s-d}, and we then discuss their features and their differences 
with the iron lines generated in Kerr spacetimes. Summary and conclusions 
are reported in Section~\ref{s-c}. Throughout the paper, we use units in which 
$G_{\rm N} = c = 1$, unless stated otherwise.

\section{K$\alpha$ iron line profile \label{s-ka}}

The X-ray spectrum of both stellar-mass and supermassive BH candidates has 
often a power-law component. This feature is commonly interpreted as the inverse 
Compton scattering of thermal photons by electrons in a hot corona above the 
accretion disk. Such a ``primary component'' irradiates also the accretion disk, 
producing a ``reflection component'' in the X-ray spectrum. The illumination of 
the cold disk by the primary component also produces spectral lines by fluorescence. 
The strongest line is the K$\alpha$ iron line at 6.4 keV. This line is intrinsically 
narrow in frequency, while the one observed appears broadened and skewed. 
The most accepted interpretation is that the line is strongly altered by special 
and general relativistic effects, which produce a characteristic profile first predicted 
in Ref.~\cite{fab89} and then observed for the first time in the ASCA data of the 
Seyfert~1 galaxy MCG-6-30-15~\cite{tan95}. For some sources, this line is 
extraordinarily stable, in spite of a substantial variability of the continuum, which 
suggests that its shape is determined by the geometry of the spacetime around 
the compact object rather than by the properties of the accretion flow.

The profile of the K$\alpha$ iron line depends on the metric of the spacetime, 
the geometry of the emitting region, the disk emissivity, and the disk's inclination 
angle with respect to the line of sight of the distant observer. The background
metric determines the propagation of the photons from the disk to the distant
observer, the velocity of the material in the accretion disk, the gravitational
redshift, and also the inner edge of the disk if the latter is supposed to be at the
ISCO radius, $r_{\rm in} = r_{\rm ISCO}$. The geometry
of the emitting regions is unknown, but the simplest option is to assume that
it ranges from the inner edge $r_{\rm in}$ to some outer radius $r_{\rm out}$.
The disk emissivity is often supposed to have a power-law behavior, 
$I_{\rm e} \propto 1/r^{\alpha}$, with an index $\alpha$ to be determined by 
fitting the data. The inclination of the disk with respect to the line of sight of 
the distant observer, say $i$, is usually another free parameter to be inferred 
during the fitting procedure. The dependence of the line profile on $a_*$,
$i$, $\alpha$, and $r_{\rm out}$ in the Kerr background has been analyzed 
in detail by many authors, starting with Ref.~\cite{fab89}. The case of iron lines
generated in other spacetimes can be found in~\cite{iron}.

In this work, we use the code described in~\cite{iron}. The photon flux number 
density measured by a distant observer is given by
\be
N_{E_{\rm obs}} &=& \frac{1}{E_{\rm obs}} 
\int I_{\rm obs}(E_{\rm obs}) d \Omega_{\rm obs} = \nonumber\\ 
&=& \frac{1}{E_{\rm obs}} \int g^3 I_{\rm e}(E_{\rm e}) 
d \Omega_{\rm obs} \, .
\ee
Here $I_{\rm obs}$ and $E_{\rm obs}$ are, respectively, the specific intensity of 
the radiation and the photon energy as measured by the distant observer, 
$d \Omega_{\rm obs}$ is the element of the solid angle subtended by the image 
of the disk on the observer's sky, $I_{\rm e}$ and $E_{\rm e}$ are, respectively, 
the local specific intensity of the radiation and the photon energy in the rest frame 
of the emitter, and $g = E_{\rm obs}/E_{\rm e}$ is the redshift factor. $I_{\rm obs} 
= g^3 I_{\rm e}$ follows from the Liouville's theorem. The disk emission is
assumed monochromatic with the rest frame energy $E_{\rm{K}\alpha} = 6.4$~keV,
and isotropic with a power-law radial profile:
\be
I_{\rm e}(E_{\rm e}) \propto \delta (E_{\rm e} - E_{\rm{K}\alpha}) / r^{\alpha} \, .
\ee
Doppler boosting, gravitational redshift, and frame dragging are encoded 
in the calculation of $g$, while the light bending enters in the integration.
More details can be found in~\cite{iron}.

\begin{figure}
\begin{center}
\includegraphics[type=pdf,ext=.pdf,read=.pdf,width=7.5cm]{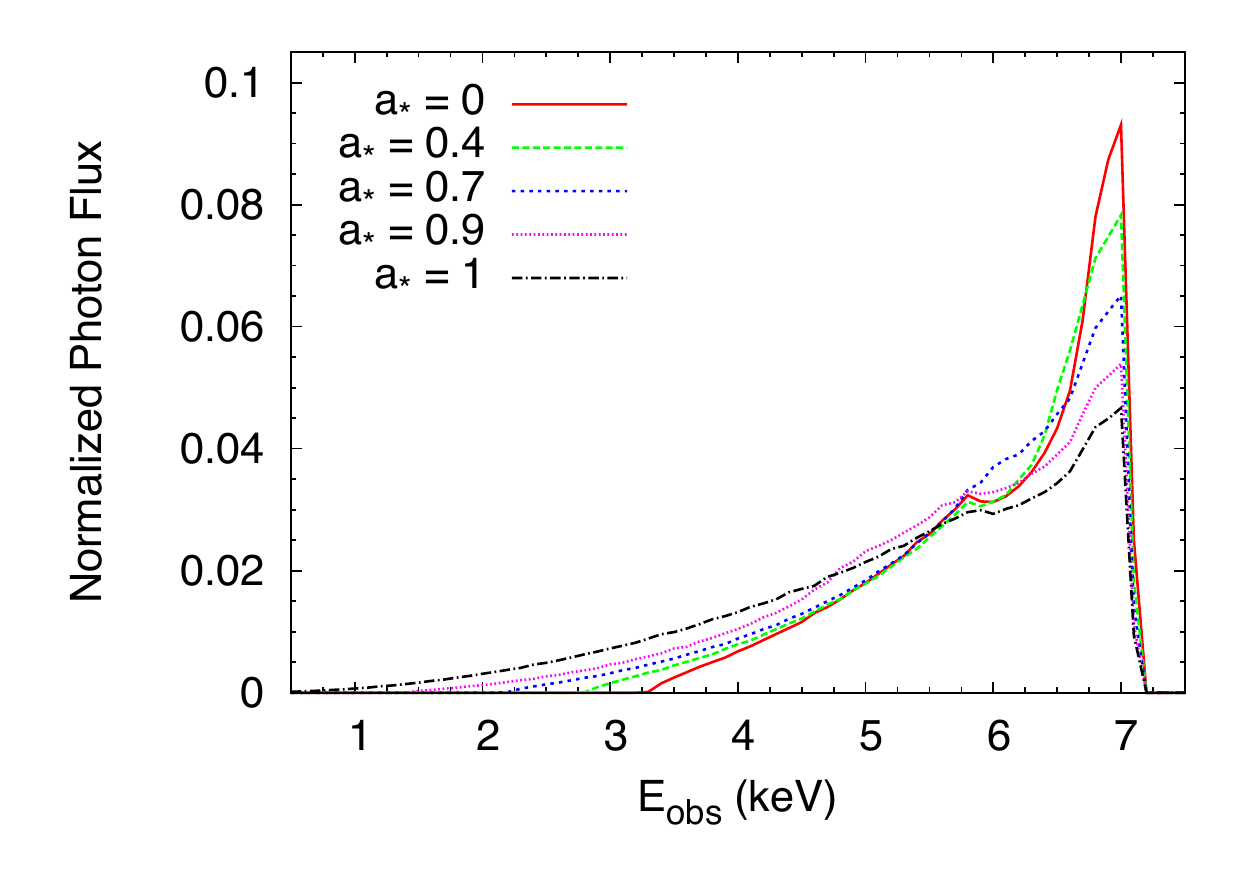}
\end{center}
\vspace{-0.7cm}
\caption{Iron line profile in the Kerr background for different values of the spin
parameter $a_*$. Model parameters: inner edge $r_{\rm in} = r_{\rm ISCO}$, 
outer edge $r_{\rm out}/M = r_{\rm in}/M + 100$, viewing angle $i = 45^\circ$, 
and index of the intensity profile $\alpha = 3$.}
\label{fig0}
\end{figure}

If we assume that astrophysical BH candidates are the Kerr BHs of general 
relativity, the analysis of the K$\alpha$ iron line can be used to estimate the
spin parameter of these objects. 
Fig.~\ref{fig0} shows the iron line profile for different values of $a_*$. 
The key-point is the assumption that the inner edge of the disk is
at the ISCO radius, which monotonically increases from $r_{\rm ISCO}/M = 6$ 
for a Schwarzschild non-rotating BH to $r_{\rm ISCO}/M = 1$ for an
extremal Kerr BH with $a_* = 1$. The iron line profile of fast-rotating Kerr BHs
is characterized by a low-energy tail, as a consequence of the strong 
gravitational redshift of the radiation emitted at small radii. Current measurements 
of the spin parameter of super-massive BH candidates with the analysis of the 
K$\alpha$ iron line (under the assumption of Kerr background) are reported in 
Tab.~\ref{tab}.

\begin{table*}
\begin{center}
\begin{tabular}{c c c c c}
\hline
\hline
AGN &  \hspace{.5cm} & $a_*$ &  \hspace{.5cm} & References \\
\hline
\hline
MGC-6-30-15 && $> 0.98$ && \cite{mcg} \\
Fairall~9 && $0.65\pm0.05$ && \cite{F9,patrick} \\
SWIFT~J2127.4+5654 && $0.6\pm0.2$ && \cite{2127} \\
1H~0707-495 && $> 0.98$ && \cite{0707} \\
Mrk~79 && $0.7\pm0.1$ && \cite{79} \\
NGC~3783 && $> 0.98$ && \cite{3783} \\
Mrk~335 && $0.70\pm0.12$ && \cite{patrick} \\
NGC~7469 && $0.69\pm0.09$ && \cite{patrick} \\
\hline
\hline
\end{tabular}
\end{center}
\vspace{-0.2cm}
\caption{Current measurements of the spin parameter of super-massive BH
candidates with the analysis of the K$\alpha$ iron line.}
\label{tab}
\end{table*}

\section{Singular and Regular spacetimes \label{s-sol}}

We shall consider here some known examples of interior solutions under the 
assumption of spherical symmetry for the spacetime. Therefore we will consider 
non-rotating interiors that match smoothly to an exterior Schwarzschild vacuum 
metric. The most general  metric describing such an interior is given in 
Schwarzschild coordinates as
\beq \label{metric}
ds^2=-e^{2\phi(r)}dt^2+\frac{1}{1-F(r)/r}dr^2+r^2d\Omega^2 \, ,
\eeq
where $d\Omega^2$ is the line element on the unit two-sphere. The energy 
momentum tensor is given by
\beq
T={\rm diag}\{\rho, p_r, p_\theta, p_\theta\} \, ,
\eeq
where the energy density $\rho$ and the radial and tangential pressures 
$p_r$ and $p_\theta$ are functions of $r$ that are related to the metric 
functions $\phi$ and $F$ via Einstein's equations. 
The function $F$ represents the gravitational mass contained within the 
radius $r$ and from the coefficient $g_{rr}$ in the metric~\eqref{metric} 
it is easy to see that at the boundary of the cloud $r_b$ we have 
$F(r_b)=2M$, where $M$ is the Schwarzschild mass.

We shall assume that the cloud is transparent to radiation and it is weakly interacting 
with the particles in an accretion disk so that viscosity is negligible and the accretion 
disk can be described by circular geodesics within the cloud. This can be the case for
a cloud of very low density or made of some exotic matter.
Then we can study the observational properties of the accretion disk that extends within 
the interior solution and determine in what respect they may differ from those  of
accretion disks around BHs. By assuming that the particles in the accretion disk follow
geodesic motion we allow in principle the disk in the inner cloud to extend all the way
to the center as the ISCO for these interiors is at $r=0$. 
For a realistic case on the other hand there would be an inner edge of the
disk $r_{\rm in}\neq 0$ that would be determined by the physical properties of the cloud.
In fact, above certain densities, viscosity and scattering effects would deviate the particles 
in the disk from circular geodesic motion. Also, since the disk is emitting light, there
will be a certain radius at which the radiation pressure of the light balances the
gravitational attraction (Eddington luminosity) thus disrupting the accretion disk.

In the following we will consider two examples for two
different kinds of interiors. Namely, we will study two interior solutions
for perfect fluids and two for a cloud sustained only by tangential pressures.
In order to distinguish the most relevant physical features we chose one regular
solution with constant density and one singular solution where the density
diverges like $r^{-2}$ for both fluid models.

As shown in \cite{dm}, we can consider these solutions as the limit 
for a slowly evolving collapsing cloud. Therefore, in the singular examples
the central singularity has to be interpreted as a region of arbitrarily high 
density where the divergence can be reached only as $t$ goes to infinity.

The study of circular geodesics and accretion disk properties in a spherically
symmetric spacetime can be done following
\cite{Thorne}.
In general, it is found that there are two different regimes of accretion
depending on where the boundary is located. For $r_b<6M$, there is a
gap between the accretion disk in the Schwarzschild sector and the accretion 
disk in the interior. For $r_b\geq 6M$, there are stable circular orbits extending
from infinity to the center. In the following we will consider always the second
accretion regime.

\subsection{Tangential pressure}

Interior solutions matching to Schwarzschild and sustained only
by tangential pressures ($p_r=0$) were first studied in 
\cite{florides}.
Einstein's equations for this case take the form
\begin{eqnarray}
  \rho &=& \frac{F'}{r^2} \, , \label{rho1} \\ 
   p_\theta &=& \frac{1}{2}\rho r \phi' \, , \label{ptheta}
    \\ 2\phi' &=& \frac{F(r)}{r^2-rF(r)} \, , \label{phi1}
\end{eqnarray}
and it is easy to see that there is the freedom to specify one
free function. In the following we choose a suitable mass
profile $F$ and then determine the other
quantities via Einstein's equations.

\subsubsection{Regular example}

To begin, we consider the constant density interior obtained by choosing
\beq
F=M_0r^3 \, .
\eeq
Then Eqs.~\eqref{rho1} and \eqref{ptheta} give
\beq
 \rho=3M_0, \; \; p_\theta=\frac{3M_0^2r^2}{4(1-M_0r^2)} \, ,
\eeq
while from Eq.~\eqref{phi1} written as $\phi'=\frac{M_0r}{2(1-M_0r^2)}$
we get
\beq
e^{2\phi}=\frac{C}{\sqrt{1-M_0r^2}} \, , 
\eeq
with the integration constant $C$ given by the matching condition 
with Schwarzschild,
\beq
C=-(1-M_0r_b^2)^{3/2} \, .
\eeq
The metric takes the form~\cite{florides}
\beq
ds^2=-\frac{(1-M_0r_b^2)^{3/2}}{\sqrt{1-M_0r^2}}dt^2
+\frac{1}{1-M_0r^2}dr^2+r^2d\Omega^2 \, ,
\eeq
and it is easy to see that there are no singularities in this spacetime. 
Furthermore the total mass for the Schwarzschild exterior is given by 
$2M=M_0r_b^3$.

\subsubsection{Singular example}

As a second example, we consider a cloud with a singularity at the center.
By choosing the free function $F$ as
\beq
F=M_0r \, ,
\eeq
from Eqs.~\eqref{rho1} and \eqref{ptheta} we get  
\beq
 \rho=\frac{M_0}{r^2} \, , \; \; p_\theta=\frac{M_0}{4(1-M_0)}\rho \, .
\eeq
In this case, the energy density diverges as $r$ goes to zero, and from the
evaluation of the Kretschmann scalar it is easy to verify that $r=0$
corresponds to a true curvature singularity.
Note that this example corresponds to a linear equation of state $\rho=k p_\theta$,
with $k=\frac{M_0}{4(1-M_0)}$.

From Eq.~\eqref{phi1}, we get $\phi'=\frac{M_0}{2(1-M_0)r}$, which, once
integrated, gives
\beq
e^{2\phi}=Cr^\frac{M_0}{1-M_0} \, , \; \; 
C=(1-M_0)r_b^{-\frac{M_0}{1-M_0} }\, ,
\eeq
\newpage
\noindent 
and the metric becomes~\cite{dm}
\beq
ds^2=-(1-M_0)\left(\frac{r}{r_b}\right)^\frac{M_0}{1-M_0}dt^2
+\frac{1}{1-M_0}dr^2+r^2d\Omega^2 \, .
\eeq

\subsection{Perfect fluid}

We now turn the attention to perfect fluid interiors that are defined 
by the relation between radial and tangential pressure as given by $p_r=p_\theta$.
These interiors have been widely studied in the literature starting from the 
pioneering work by Tolman~\cite{tolman}. 
Einstein's equations take the form
\begin{eqnarray}
  \rho &=& \frac{F'}{r^2} \, ,\label{rho2} \\ 
   p &=&  \frac{2\phi'}{r}\left[1-\frac{F(r)}{r}\right]-\frac{F(r)}{r^3} \, , \label{p}
    \\ p' &=& -(\rho+p)\phi' \, , \label{phi2}
\end{eqnarray}
and once again the system is completely determined once we specify the 
mass function $F$. Nevertheless, in this case, solving the system of Einstein's 
equations is more difficult since after the choice of $F$ we can substitute 
$\phi'$ from Eq.~\eqref{p} into Eq.~\eqref{phi2} in order to obtain the differential 
equation that must be satisfied by $p$. This is known as the 
Tolman-Oppenheimer-Volkoff (TOV) equation
\begin{equation}\label{TOV}
    p'=-(\rho+p)\frac{\left[pr^3+F(r)\right]}
{2r\left[r-F(r)\right]} \, ,
\end{equation}
and the system is solved once a solution for Eq.~\eqref{TOV} is found. This is 
possible in some very simple special cases like the ones provided below.
For a comprehensive list of known solutions for perfect fluid interiors, see 
for example~\cite{DL}.

\subsubsection{Regular example}

The simplest model is the constant density Schwarzschild interior with
\begin{equation}
    F(r)=M_0r^3 \, .
\end{equation}
Then again $\rho=3M_0$ and solving the TOV equation gives the pressure
\beq
p=3M_0\frac{\sqrt{1-M_0r_b^2}-
\sqrt{1-M_0r^2}}{\sqrt{1-M_0r^2}-3\sqrt{1-M_0r_b^2}} \, ,
\eeq
which vanishes at the boundary $r_b$ where the cloud matches the Schwarzschild 
exterior with total mass $M=M_0r_b^3$. Then from Eq.~\eqref{p} we get
\beq
e^{2\phi}=(A-\sqrt{1-M_0r^2})^2 \, ,
\eeq
with the integration constant $A$ given by the matching condition
\beq
A=2\sqrt{1-M_0r_b^2} \, .
\eeq
The metric then takes the form~\cite{buchdahl}
\beq
ds^2=-(A-\sqrt{1-M_0r^2})^2dt^2+\frac{1}{1-M_0r^2}dr^2+r^2d\Omega^2 \, .
\eeq
Note that from the pressure profile we can obtain a condition for the pressure to
be positive and finite at the center $p(0)>0$. This implies the existence of a lower
boundary limit $r_b^2=\frac{8}{9M_0}$ at which the central pressure diverges. 
This corresponds to imposing a minimum boundary greater than the
Schwarzschild radius as $r_b>\frac{9M}{4}$.

\subsubsection{Singular example}

As we did in the tangential pressure case, we now investigate a perfect fluid 
interior where the density diverges at the center. This is again given by the choice
\begin{equation}
    F(r)=M_0r \, ,
\end{equation}
for which
\begin{equation}
    \rho=\frac{M_0}{r^2} \, .
\end{equation}
To solve the TOV Eq.~\eqref{TOV}, we must define a new parameter $\lambda$ as
\begin{equation}
    \lambda=\sqrt{\frac{1-2M_0}{1-M_0}} \, , \qquad
M_0 = \frac{1-\lambda^2}{2-\lambda^2} \, .
\end{equation}
So that once we integrate the TOV equation we get the pressure as
\begin{equation}
    p = \frac{1}{2-\lambda^2}\frac{1}{r^2}
\left(\frac{(1-\lambda)^2A-(1+\lambda)^2Br^{2\lambda}}{A-Br^{2\lambda}}\right) \, ,
\end{equation}
where $A$ and $B$ are constants coming from the integration.
From the integration of Eq.~\eqref{p}, we get 
\begin{eqnarray}
  e^{2\phi} = (Ar^{1-\lambda}-Br^{1+\lambda})^2 \, .
\end{eqnarray}
The two integration constants $A$ and $B$ are determined by the matching 
condition for $g_{tt}$ and by the condition that $p$ must vanish at the boundary.
They can be written as
\begin{eqnarray}
    A&=& \frac{(1+\lambda)^2r_b^{\lambda-1}}{4\lambda\sqrt{2-\lambda^2}} \, ,  \\
    B&=& \frac{(1-\lambda)^2r_b^{-\lambda-1}}{4\lambda\sqrt{2-\lambda^2}} \, ,
\end{eqnarray}
and the metric takes the form~\cite{dm}
\begin{equation}\label{metric-sol}
    ds^2=-(Ar^{1-\lambda}-Br^{1+\lambda})^2dt^2+(2-\lambda^2)dr^2
+r^2d\Omega^2 \, .
\end{equation}
Note that in this case the relation between $\rho$ and $p$ is more complicated than in
the corresponding tangential pressure case, but it tends to a stiff fluid equation of state 
$\rho=p$ in the limit $\lambda=0$ (which corresponds to $M_0=1/2$) and thus implies
a minimum boundary radius of $r_b=4M$, greater than the one obtained for the 
regular example above.

\section{Results and discussion \label{s-d}}

We can now use the solutions presented in the previous section to compute
the associated K$\alpha$ iron line. The free parameters of the model are:
the radius of the compact object, $r_b$, the inner and outer edges of the 
accretion disk, $r_{\rm in}$ and $r_{\rm out}$, the disk's inclination angle with
respect to the line of sight of the distant observer, $i$, and the index of the
intensity profile, $\alpha$. The dependence of the iron line on $r_{\rm out}$,
$i$, and $\alpha$ is the same of the one in the Kerr geometry and it is already
well studied in the literature. So, in what follows we will always consider
the specific case with $r_{\rm out} = r_{\rm in} + 100 \, M$, $i = 45^\circ$, and
$\alpha = 3$. We will thus focus our attention on the effects of $r_b$ and 
$r_{\rm in}$.

In the Kerr background, $r_{\rm in}$ is supposed to be the ISCO radius. 
For the interior solutions described in the previous section, there is no ISCO
radius, in the sense that circular orbits are always stable. The reason
is that gravity is never so strong to make the orbit unstable, and this is due 
to the fact that the mass of the objects is not concentrated at $r=0$, and the
total mass inside the sphere of radius $r$ decreases as $r$ decreases.
As said, the choice $r_{\rm in} = 0$ is likely not very physical, because the density 
of the disk at small radii would diverge, with the result that the thin disk 
approximation would break down. We can assume $r_{\rm in}$ a free 
parameter, but we will see that our conclusions are not sensitive to the exact 
choice of $r_{\rm in}$. 
As $r_{\rm ISCO} = 6 \; M$ in the Schwarzschild background, to prevent
the formation of two disconnected disks with strange features (see the
discussion in the previous section and, for more details, the one 
in Ref.~\cite{dm}), we will assume that the physical radius of 
the compact object is $r_b \ge 6 \; M$.

\begin{figure*}
\begin{center}
\includegraphics[type=pdf,ext=.pdf,read=.pdf,width=7.5cm]{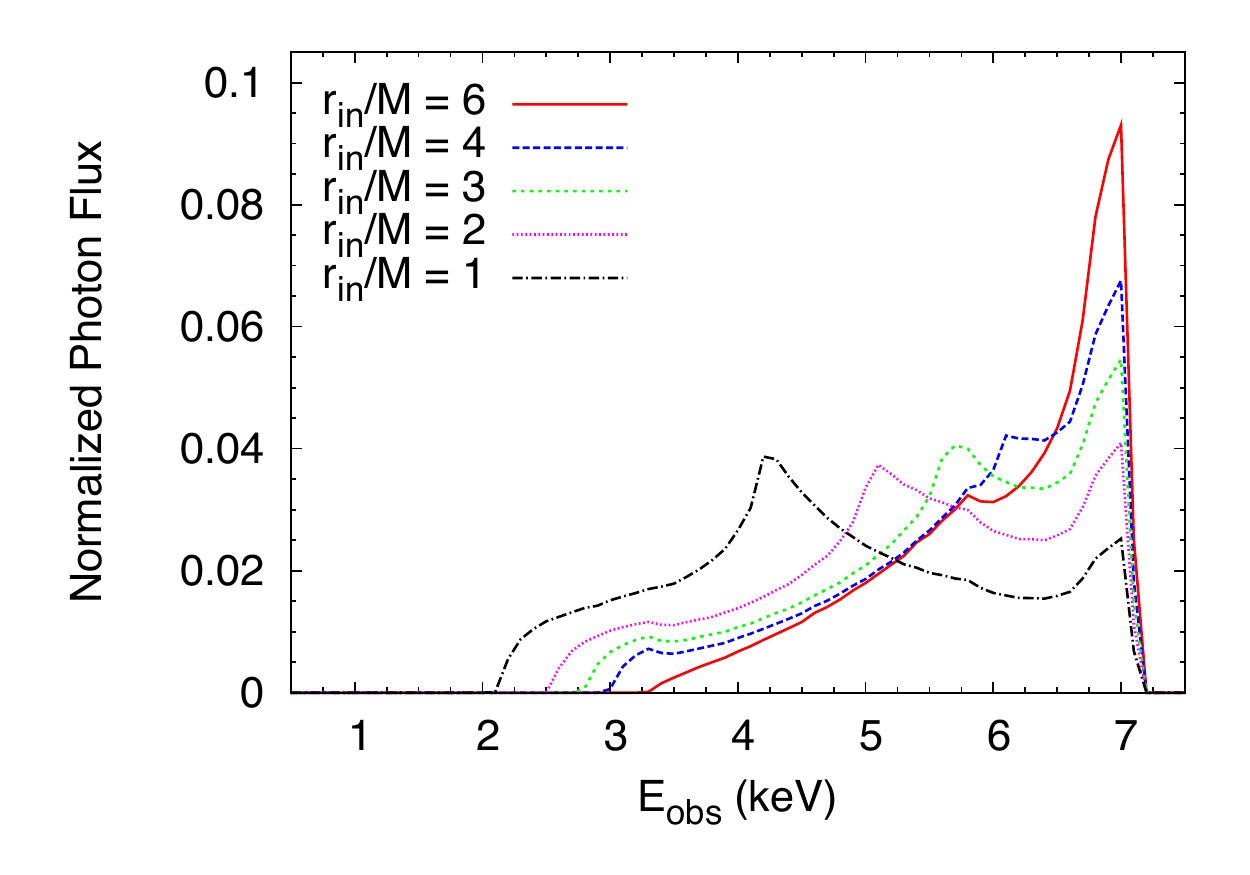}
\hspace{0.5cm}
\includegraphics[type=pdf,ext=.pdf,read=.pdf,width=7.5cm]{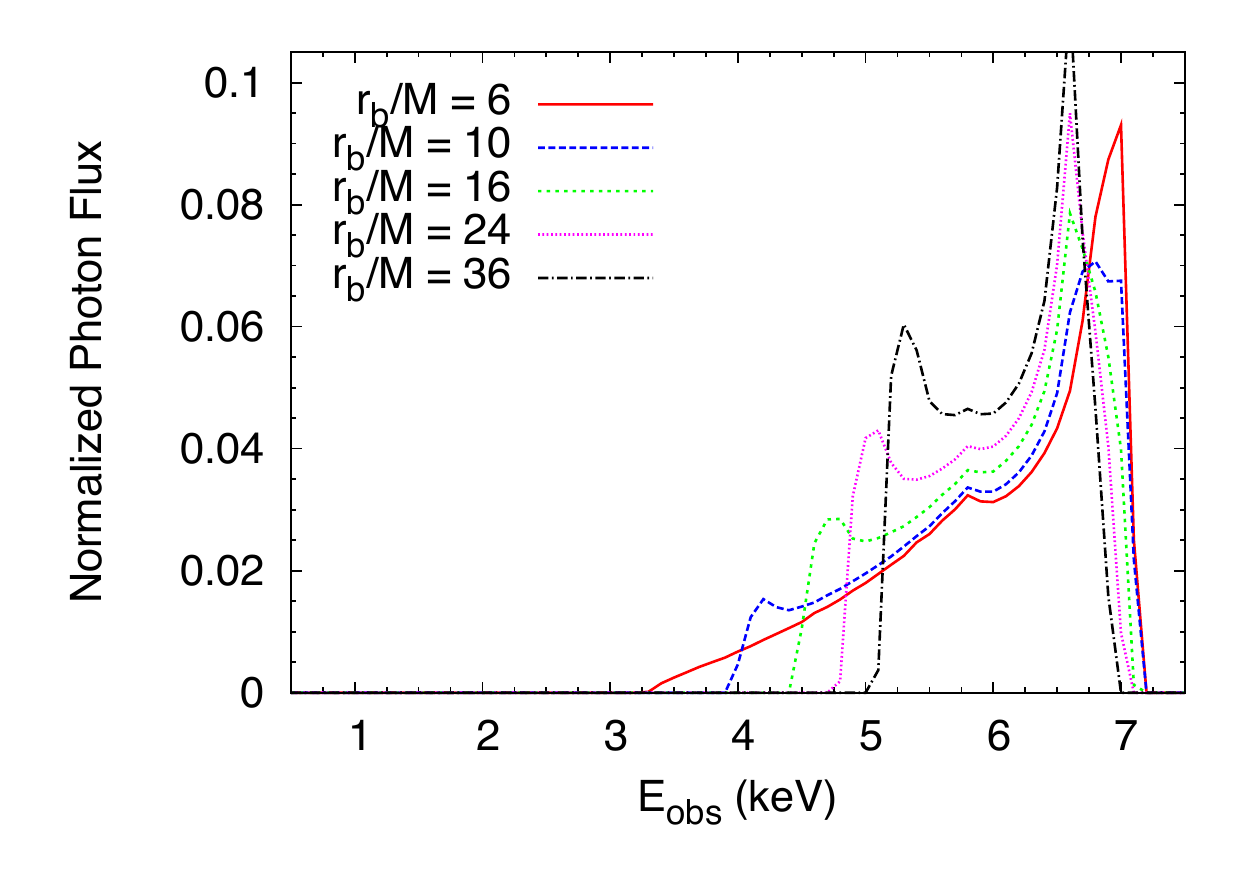}
\end{center}
\vspace{-0.7cm}
\caption{Iron line profile in the singular tangential pressure case. 
Left panel: $r_{\rm b}/M = 6$ and $r_{\rm in}/M = 6$, 4, 3, 2, and 1. 
Right panel: $r_{\rm in}/M = 6$ and $r_{\rm b}/M = 6$, 10, 16, 24, and 36. 
The outer edge is set at $r_{\rm out}/M = r_{\rm in}/M + 100$, the viewing 
angle is $i = 45^\circ$, and the index of the intensity profile is $\alpha = 3$.}
\label{fig2}
\end{figure*}

\begin{figure*}
\begin{center}
\includegraphics[type=pdf,ext=.pdf,read=.pdf,width=7.5cm]{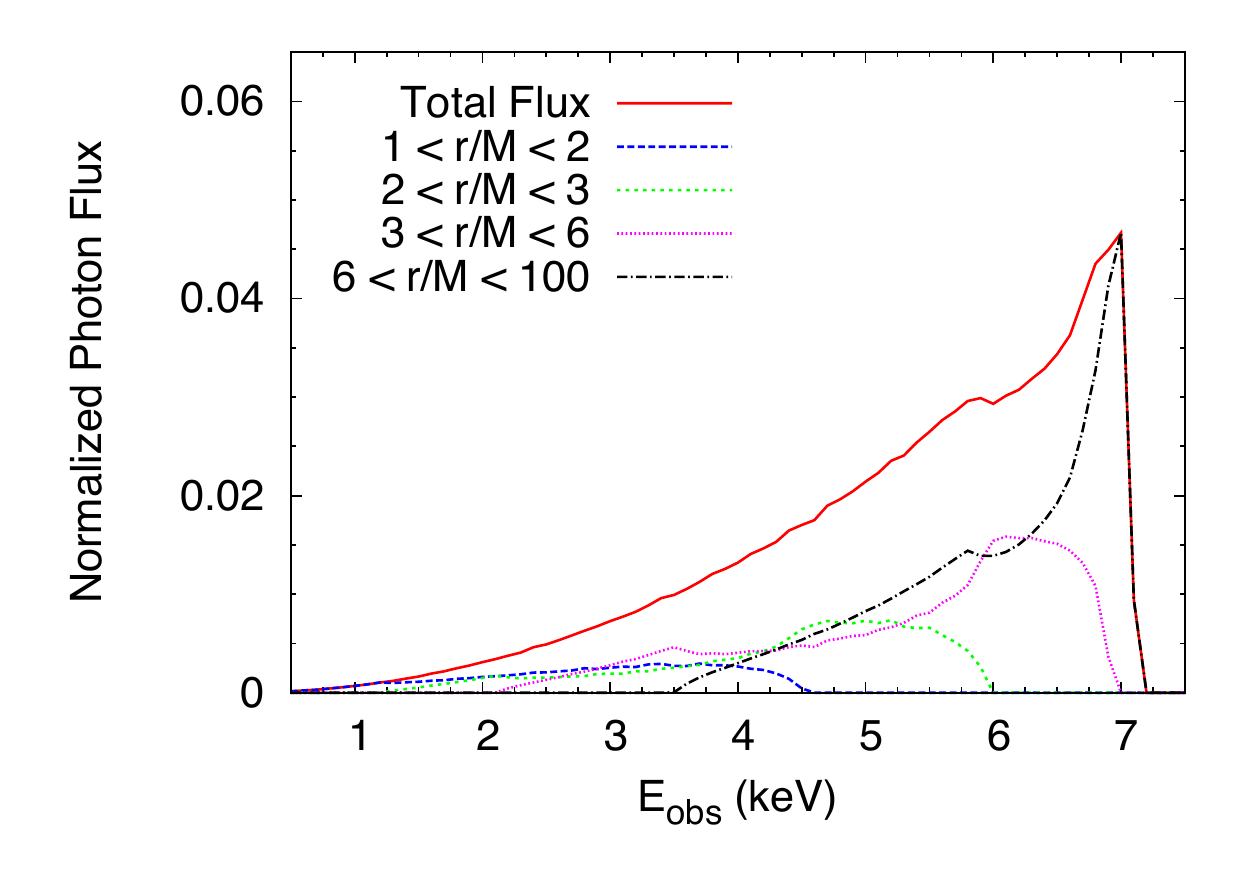}
\hspace{0.5cm}
\includegraphics[type=pdf,ext=.pdf,read=.pdf,width=7.5cm]{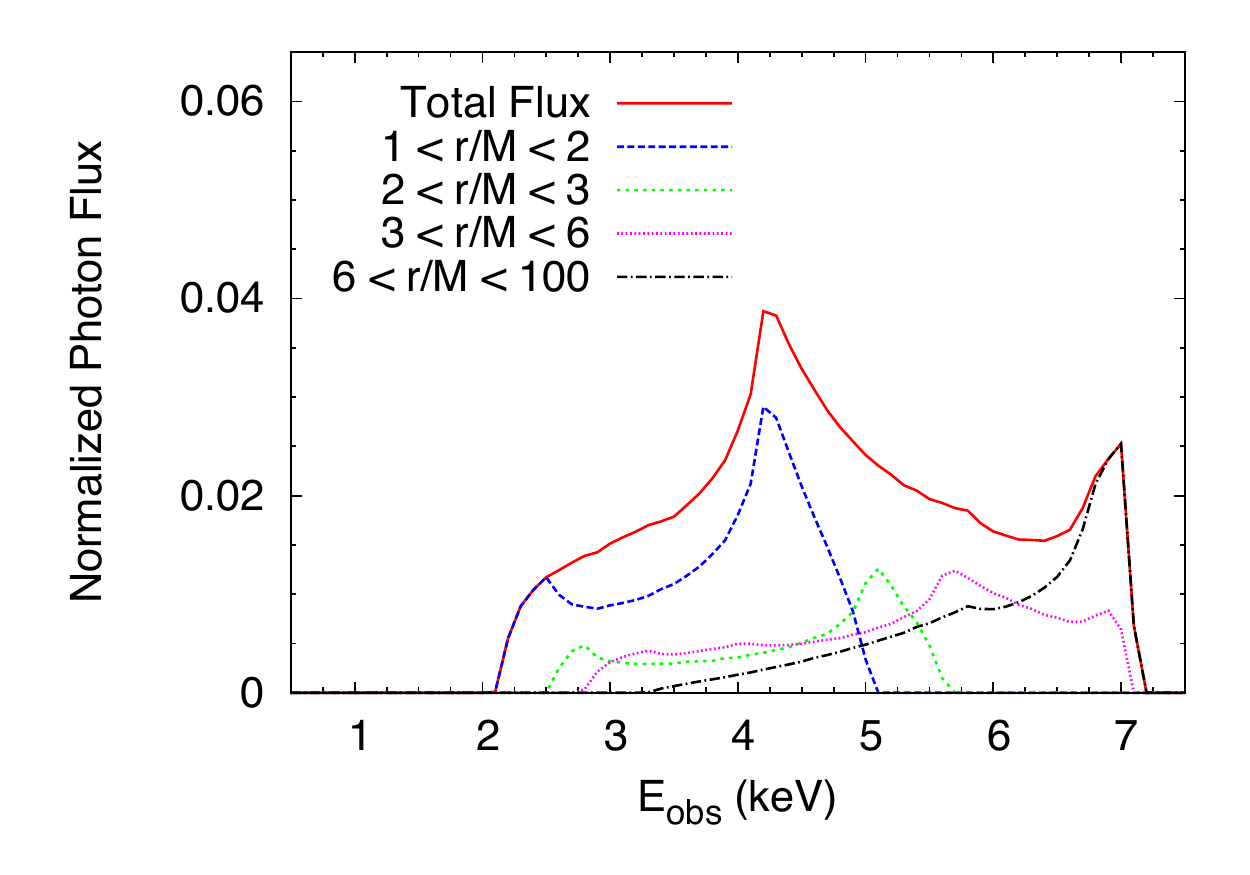}
\end{center}
\vspace{-0.7cm}
\caption{Iron line profile in the case of Kerr background with $a_* = 1$ (left panel)
and singular tangential pressure metric (right panel). 
In both examples, the inner edge of the disk
is supposed at the radial coordinate $r_{\rm in}/M = 1$, the outer edge is set at 
$r_{\rm out}/M = 100$, the viewing angle is $i = 45^\circ$, and the index of the
intensity profile is $\alpha = 3$. The panels show the total flux as well as the
corresponding contributions from the disk regions $1 < r/M < 2$, $2 < r/M < 3$,
$3 < r/M < 6$, and $6 < r/M < 100$.}
\label{fig1}
\end{figure*}

Let us start considering the singular tangential pressure case. If we set
$r_{\rm in} = r_b = 6 \; M$, we find an iron line very similar to the one from
a Schwarzschild BH. The small difference is due to the fact that around a
Schwarzschild BH some photons may cross the photon capture sphere 
and be swallowed by the BH (the effect is larger for edge-on disks). 
If we reduce the value of $r_{\rm in}$ without changing the one of $r_b$, 
we get the profiles shown in the left panel of Fig.~\ref{fig2}. On the other
hand, if we increase the value of $r_b$ maintaining $r_{\rm in} = 6 \; M$,
we find the profiles in the right panel of Fig.~\ref{fig2}. The resulting iron 
line for small values of $r_{\rm in}$ can be better understood if we compare
it with the iron line generated around an extremal Kerr BH, paying 
attention to the contribution from different radii. This is done in Fig.~\ref{fig1}.
In both the models, $r_{\rm in} = M$. There are clearly some important
differences. First, there are many photons coming from the region
$1 < r/M < 2$ of the exotic compact object, while the photon contribution
from the same region is very low in the Kerr BH case. In the former case,
the contribution is important because the intensity goes like $1/r^3$.
However, in the BH case many photons emitted from the region close to the
compact object cannot reach the distant observer because swallowed by 
the BH. In other words, the effect of light bending at small radii is quite
different in the two backgrounds. Second, a remarkable difference between
the two iron line profiles is the low-energy tail. In the case of the extremal 
BH, the photons coming from the region at small radii are strongly 
gravitational redshifted. This is simply the result of the presence of the 
event horizon -- at the event horizon, the gravitational redshift is zero.
In the case of the exotic compact object, there is no event horizon and 
therefore the gravitational redshift is much weaker. At large radii,
$6 < r/M < 100$, the two models have roughly the same line (the difference
in the left and right panels is just the normalization, but the shape is 
practically the same). The difference in the iron line profiles shown in
the right panel of Fig.~\ref{fig2} is mainly due to the different gravitational 
redshift and to the different angular velocity of the accreting gas.
The latter is smaller in the case of exotic compact object, so the Doppler
redshift and blueshift are milder. For instance, the characteristic peak
at high energies in the profile is due to the Doppler blueshift at relatively
large radii ($r/M \sim 10$), where the gravitational redshift is subdominant.
When $r_b/M \gtrsim 10$, the peak moves to lower energies.

\begin{figure*}
\begin{center}
\includegraphics[type=pdf,ext=.pdf,read=.pdf,width=7.5cm]{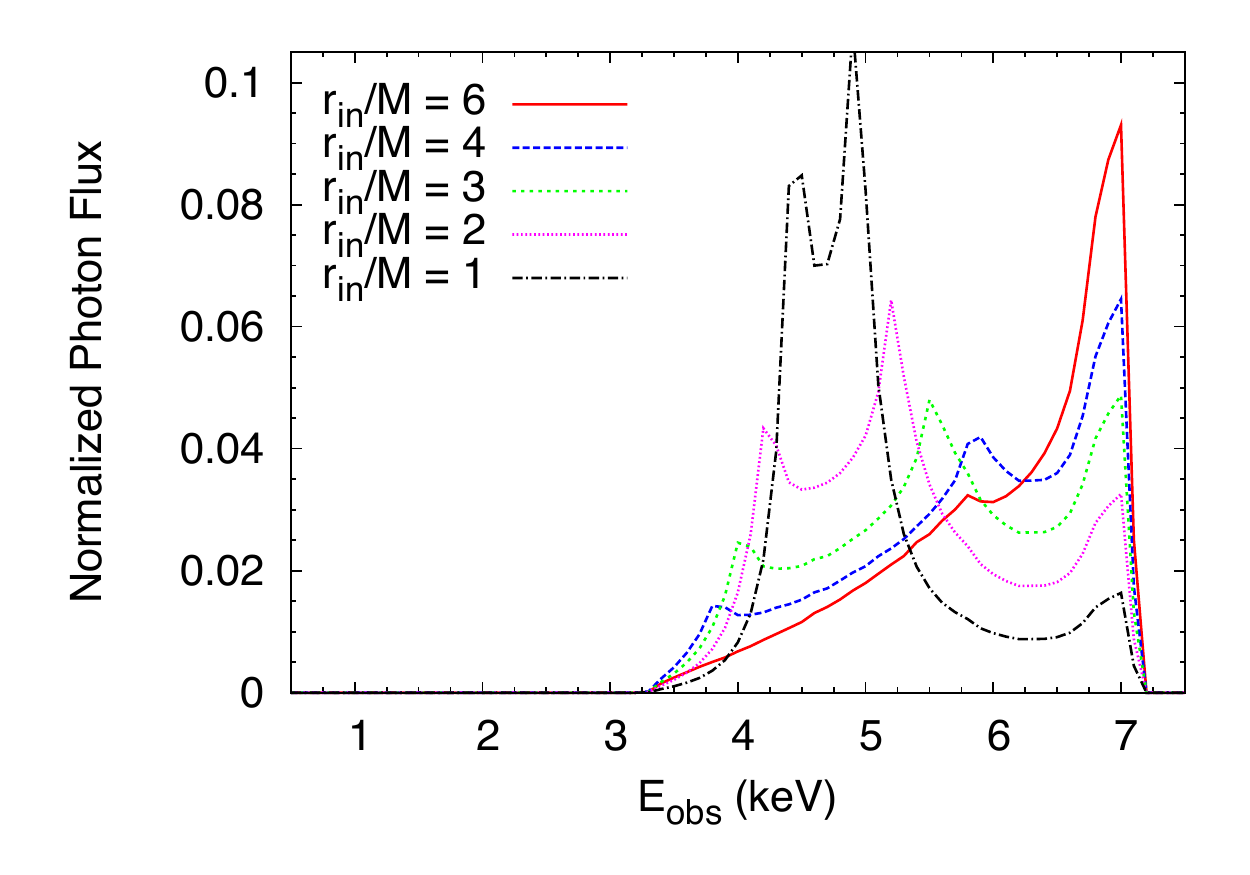}
\hspace{0.5cm}
\includegraphics[type=pdf,ext=.pdf,read=.pdf,width=7.5cm]{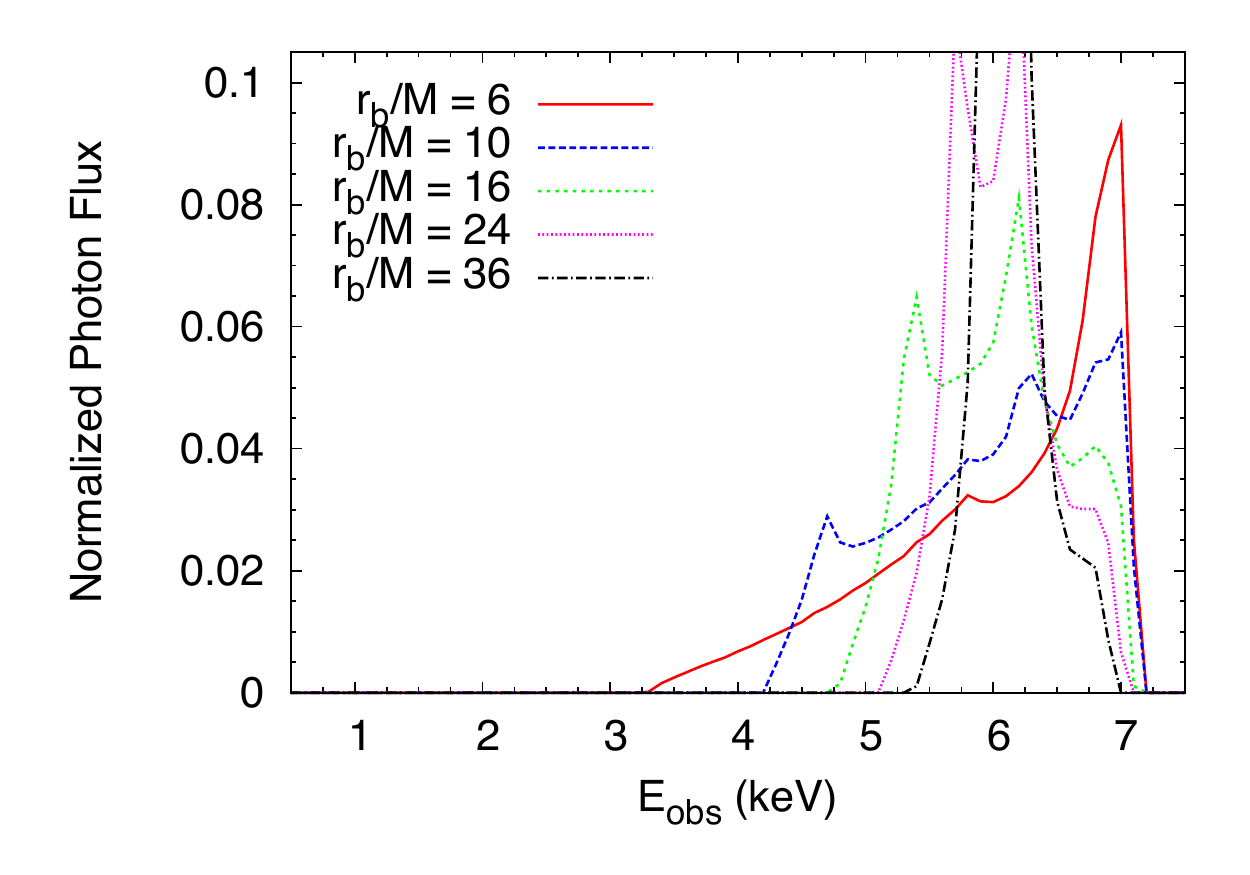}
\end{center}
\vspace{-0.7cm}
\caption{As in Fig.~\ref{fig2} for the regular tangential pressure metric.}
\label{fig4}
\end{figure*}

\begin{figure*}
\begin{center}
\includegraphics[type=pdf,ext=.pdf,read=.pdf,width=7.5cm]{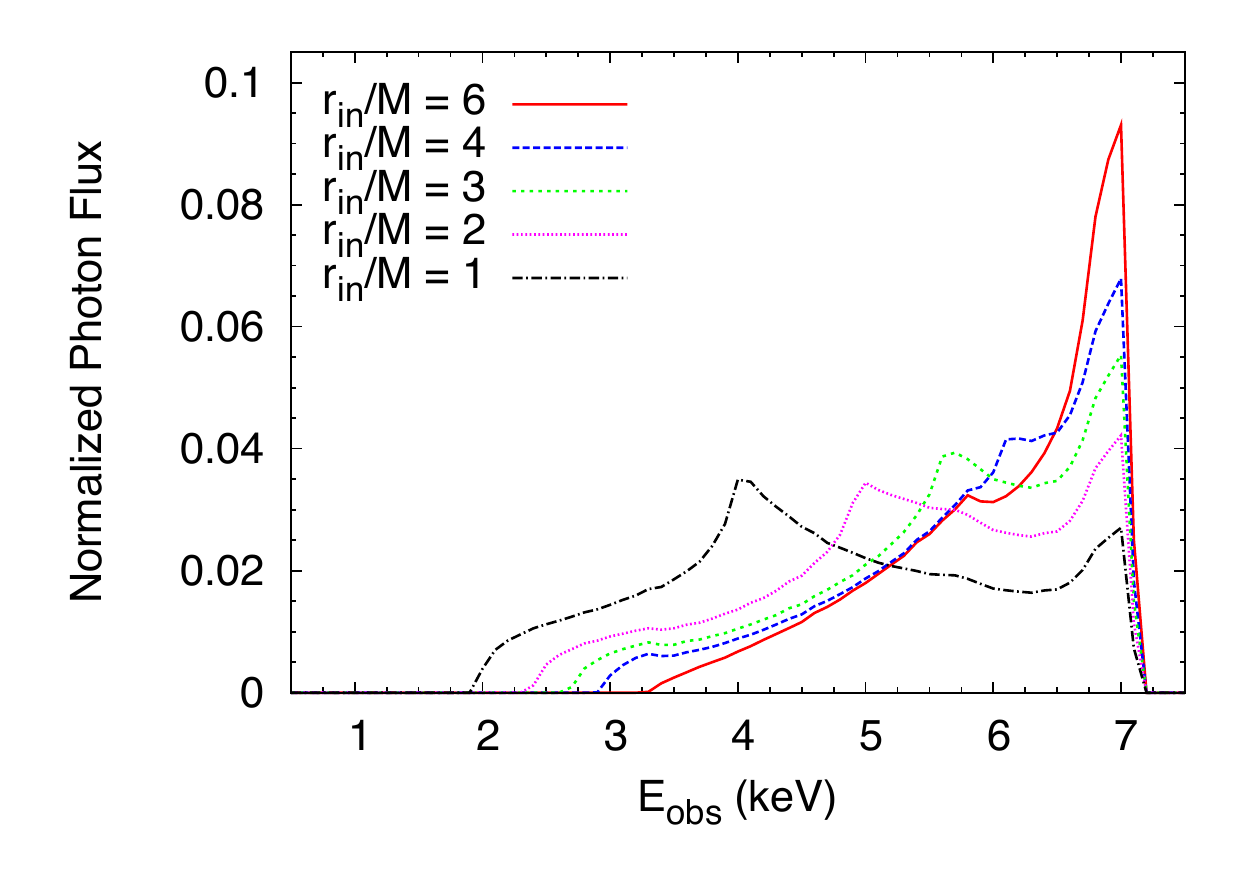}
\hspace{0.5cm}
\includegraphics[type=pdf,ext=.pdf,read=.pdf,width=7.5cm]{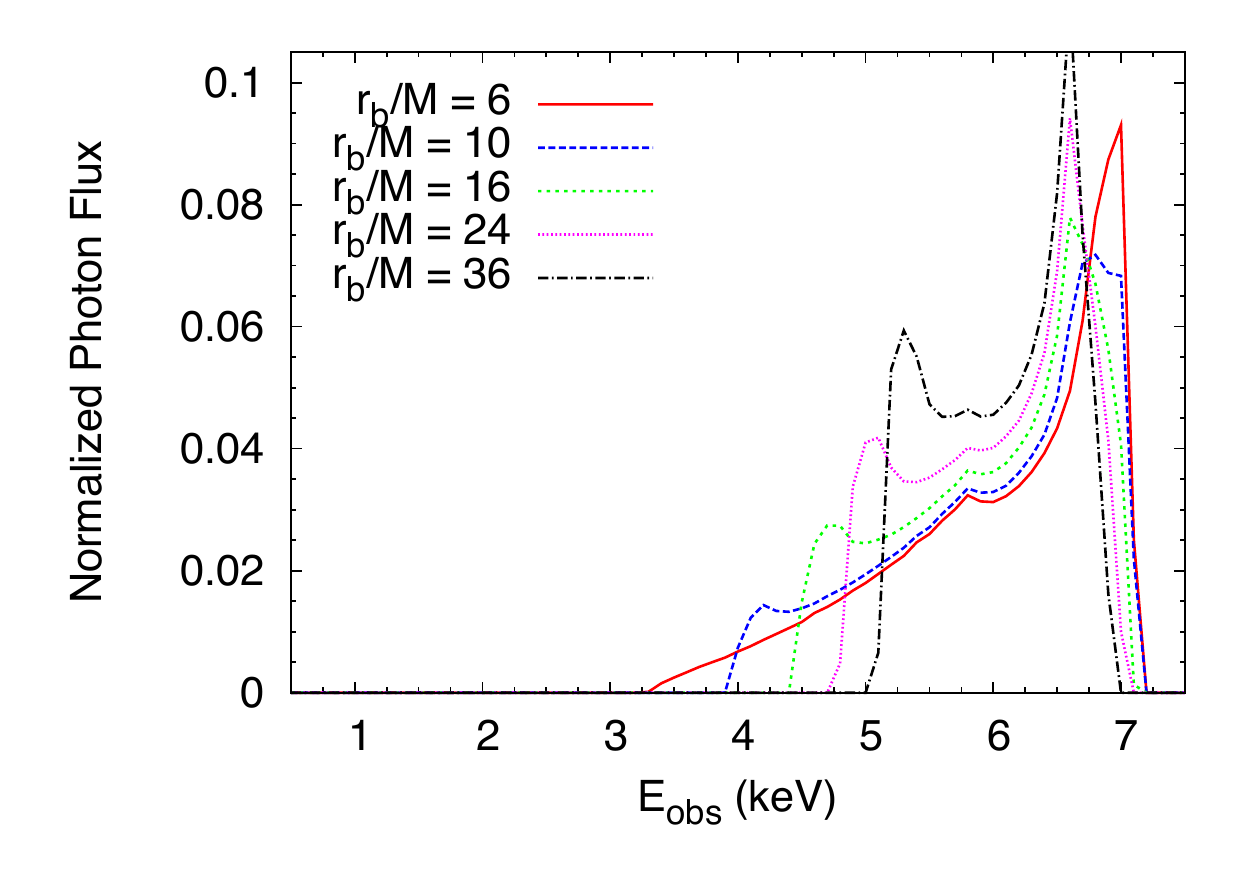}
\end{center}
\vspace{-0.7cm}
\caption{As in Fig.~\ref{fig2} for the singular perfect fluid metric.}
\label{fig6}
\end{figure*}

\begin{figure*}
\begin{center}
\includegraphics[type=pdf,ext=.pdf,read=.pdf,width=7.5cm]{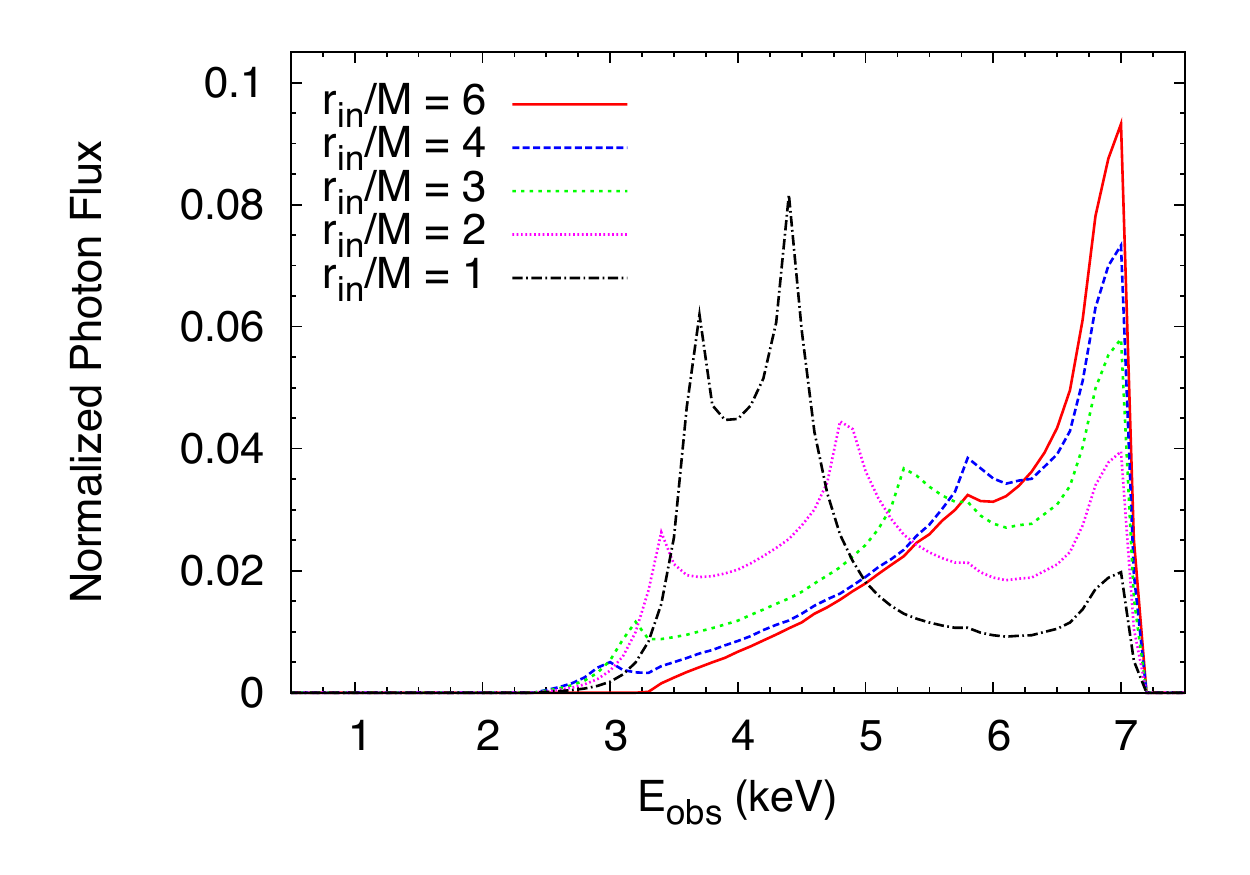}
\hspace{0.5cm}
\includegraphics[type=pdf,ext=.pdf,read=.pdf,width=7.5cm]{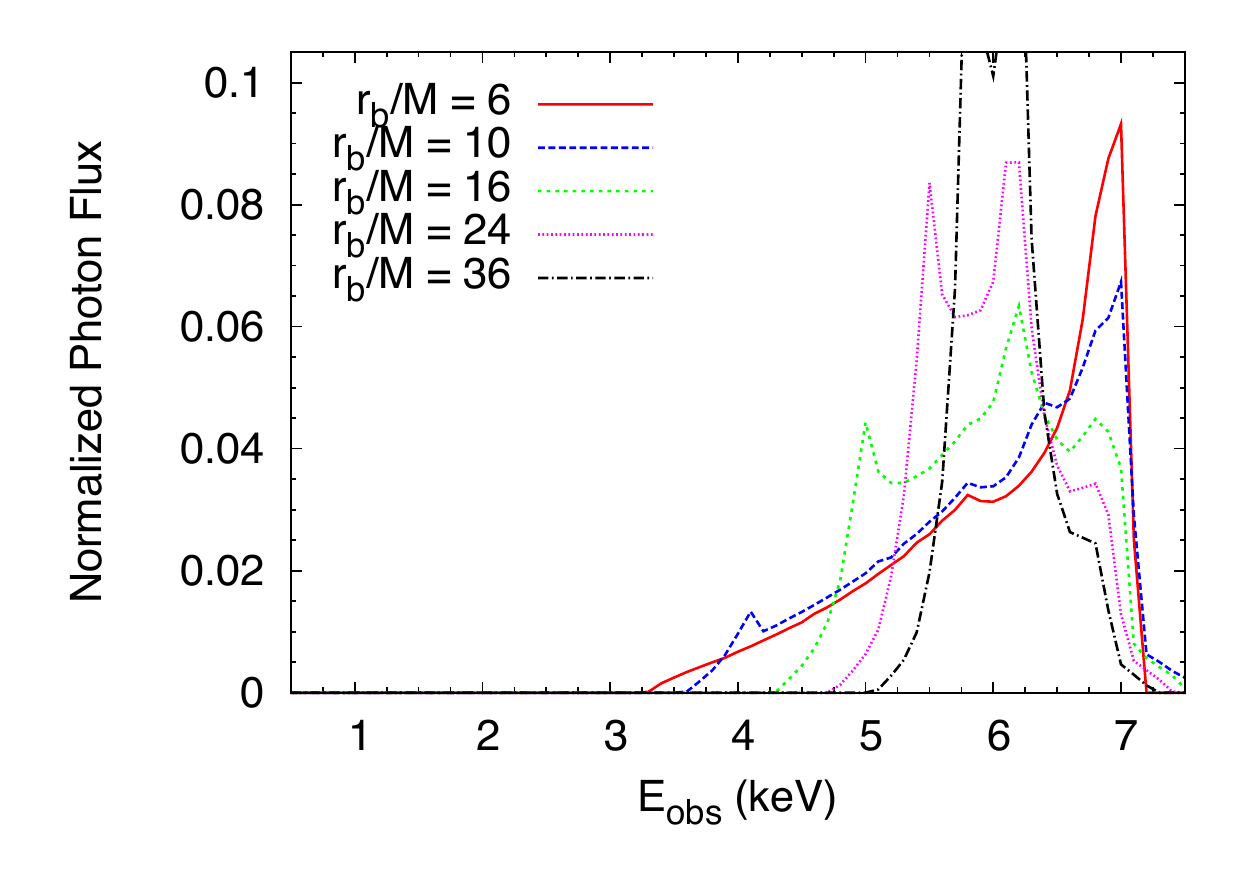}
\end{center}
\vspace{-0.7cm}
\caption{As in Fig.~\ref{fig2} for the regular perfect fluid metric.}
\label{fig7}
\end{figure*}

We can now repeat the calculations for the other interior solutions. The
qualitative features of the new iron line profiles are more or less the same
as the singular tangential pressure case. Fig.~\ref{fig4} shows
the regular tangential pressure scenario. We can find the same
characteristic features. The gravitational redshift is not as important as
in the BH background even when the inner edge of the disk is at small
radii. The effect of light bending is also an important difference with the
BH case, and photons emitted at small radii can easily reach the 
distant observer, with the clear result that their contribution in the line
profile is quite significant. The angular frequency of the gas in the accretion
disk is lower than the one around a BH, so the Doppler redshift and 
blueshift are milder. The iron lines of the perfect fluid solutions are
reported in Figs.~\ref{fig6} (singular case) and \ref{fig7} (regular case).

\section{Summary and conclusions \label{s-c}}

The super-massive objects at the center of galaxies are thought
to be the Kerr BHs of general relativity, but their actual nature has still to be
verified. In the present paper, we have explored the possibility that these
objects are instead exotic compact bodies described by a so called ``interior'' 
solution that smoothly matches to a known vacuum exterior, and we have
looked for possible observational signatures to observationally distinguish
this class of objects from Kerr BHs. Today, the only available
technique to probe the spacetime geometry around super-massive BH
candidates is the analysis of the K$\alpha$ iron line. We have thus
calculated the iron lines that can be generated in these spacetimes and
then compared with the ones of Kerr BHs. It turns out that the profiles of
the iron line of this class of exotic objects have some specific observational
signatures. In particular:
\begin{enumerate}
\item Despite that the inner edge of the accretion disk can be at much 
smaller radii than the one of the disk around a Kerr BH (a key-feature of 
these solutions is the absence of a finite ISCO radius), the iron line cannot 
have the characteristic low-energy tail expected for the line of a fast-rotating
BH. These is a consequence of the mild gravitational redshift.
\item Photons emitted by the inner part of the accretion disk, close to the
compact object, can more easily escape to infinity and reach the distant
observer. The contribution of the radiation emitted at small radii is thus
significant, with the effect of producing a peak in the profile absent in the
BH case. Photons emitted at small radii around a Kerr BH are instead
easily captured by the latter, so their contribution to the total photon flux
detected at infinity is small.
\item If the physical radius of the compact object $r_b$ exceeds
$10 - 15$ gravitational radii, the maximum Doppler blueshift is reduced,
so the high energy peak in the iron line profile moves to lower energies
with respect to the BH case.
\end{enumerate}

The general features described above are common to all the four ``interiors''
studied here and we can argue that they are most likely common to any
reasonable interior model.
Therefore even if here we have only considered some {\it ad hoc} cases of 
non-rotating objects, we stress that our conclusions are likely to remain 
true even in the more realistic case of rotating bodies. In the BH case, the spin of
the compact object simply moves the inner edge of the disk to smaller 
radii, so that the line of fast-rotating Kerr BHs is characterized by a low-energy
tail due to the strong gravitational redshift at small radii. In the case of
our exotic compact objects, there is no ISCO and therefore a non-vanishing
rotation only introduces small corrections in the photon propagation and
in the angular velocity of the gas in the accretion disk. We thus expect that 
the associated iron line is quite similar to the one of the non-rotating case
computed in this work.

In conclusion, the K$\alpha$ iron line of the class of exotic objects considered
in this paper has specific features to be observationally distinguished from the
one associated to a Kerr BH. The few super-massive BH candidates whose iron
line has been analyzed in detail in the literature do not have an iron line with
these features and therefore we can exclude the possibility that they are
the `interior' solutions studied in this paper.
However, we are talking about 6~sources (see Tab.~\ref{tab}), 
while the super-massive BH 
candidates in the visible Universe should be around $10^{12}$, and it is not
impossible that some of them are exotic compact bodies whose nature 
differs significantly from that of BHs. Therefore the possibility remains that
some of these candidates can be described by an interior solution 
similar to the ones presented here, either regular or singular.

\begin{acknowledgments}
This work was supported by the Thousand Young Talents Program 
and Fudan University.
\end{acknowledgments}


\end{document}